\documentclass[a4paper]{article}

\usepackage{INTERSPEECH2019}

\title{Quantifying Cochlear Implant Users' Ability for Speaker Identification
using CI Auditory Stimuli 
}
\name{Nursadul Mamun, Ria Ghosh, John H. L. Hansen}

 \address{
  CRSS: Center for Robust Speech Systems; Cochlear Implant Processing Laboratory (CILab) Department of Electrical and Computer Engineering, University of Texas at Dallas, USA
  }
\email{(nursadul.mamun, ria.ghosh, john.hansen)@utdallas.edu}

\begin{document}
\pagenumbering{roman}
\maketitle
\begin{abstract}
  Speaker recognition is a biometric modality that uses underlying speech information to determine the identity of the speaker. Speaker Identification (SID) under noisy conditions is one of the challenging topics in the field of speech processing, specifically when it comes to individuals with cochlear implants (CI). This study analyzes and quantifies the ability of CI-users to perform speaker identification based on direct electric auditory stimuli. CI users employ a limited number of frequency bands (8 $\sim$ 22) and use electrodes to directly stimulate the Basilar Membrane/Cochlear in order to recognize the speech signal. The sparsity of electric stimulation within the CI frequency range is a prime reason for loss in human speech recognition, as well as SID performance. Therefore, it is assumed that CI-users might be unable to recognize and distinguish a speaker given dependent information such as formant frequencies, pitch etc. which are lost to un-simulated electrodes. To quantify this assumption, the input speech signal is processed using a CI Advanced Combined Encoder (ACE) signal processing strategy to construct the CI auditory electrodogram. The proposed study uses 50 speakers from each of three different databases for training the system using two different classifiers under quiet, and tested under both quiet and noisy conditions. The objective result shows that, the CI users can effectively identify a limited number of speakers. However, their performance decreases when more speakers are added in the system, as well as when noisy conditions are introduced. This information could therefore be used for improving CI-user signal processing techniques to improve human SID. 

\end{abstract}

\section{Introduction}

A cochlear implant is an implantable electric device that allows people with sensorineural hearing loss to recover hearing abilities especially speech recognition. Efficient encoding of temporal information in current CI signal processing strategy allows most of the CI users to achieve over 80\% speech understanding in quite acoustic condition \cite{rouger2007evidence, ali2018cci}. However, other aspects of auditory processing (such as speech understanding in noise, distinguishing speaker’s identity, gender, emotion) remain difficult for CI users to interpret, which is important for a better life \cite{munson2005phonetic, mamun2015prediction, gideon2017progressive}. Those difficulties could be due to the limited access of frequency channels involve with their electrodes than that of normal hearing person. Research also shows that, formant frequencies or spectral peaks of human voice which are critical for speech recognition also reflect the individual anatomical and physiological properties and thus carry the information of speaker’s identity \cite{fellowes1997perceiving, yousefi2018assessing}.  

The cochlear implant provides the sense of sound by directly stimulating the auditory nerve. The implant comprises of two important components coupled using a powerful magnet- first is an external sound processor and the other is a surgically implanted electrode array (16-22 electrode long) connecting to the auditory nerve. At a time only, a limited number of electrodes can be stimulated in the electrode-array, based on the mechanism and design of the array implanted \cite{embc2019}.
\def\factor{0.9}
\begin{figure}[t]
  \includegraphics[width=\factor\linewidth]{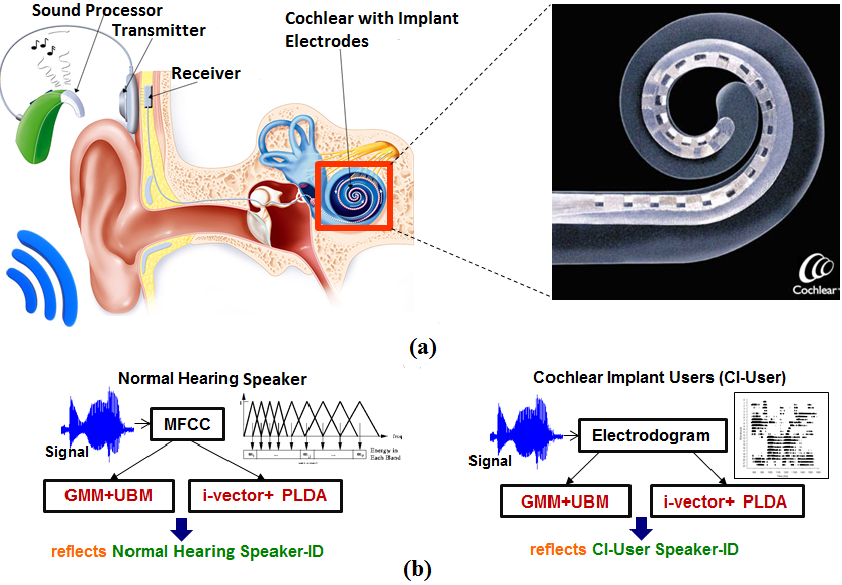}
  \caption{(a) Cross-section of the human ear high-lighting the implanted electrodes in the cochlea. (b) Comparison of the SID mechanism between Normal Hearing Person and CI-user.}
  \vspace{-10pt}
     \label{fig:Basic_blk}
 \end{figure}

Figure~\ref{fig:Basic_blk} illustrates the basic speaker identification approaches for normal hearing subjects and CI users. As shown in Fig.~\ref{fig:Basic_blk} speaker identification ability for normal hearing subjects are quantified by extracting Mel Frequency Cepstral Coefficients (MFCC) features from the speech signal and using Gaussian mixture model (GMM) or probabilistic linear discriminant analysis (PLDA) as classifier. Instead of using MFCC features, CI user’s performance in speaker identification is quantified by extracting electrodograms from speech signal using CI processor. A normal hearing person can perform efficient speaker recognition when the individual-speaker features are as widely separated as possible \cite{hansen2015speaker,khalil2019robust, chowdhury2017text, xia2019cross, islam2016robust}. Based on this feature extraction mechanism, MFCC have been widely used in automated speaker recognition system as an attempt to mimic the speaker recognition capabilities of humans. The MFCC algorithm has a Mel filter Bank that comprises of 40 Mel filters which are used to generate the MFCC coefficients and collect the speaker-related information effectively \cite{tiwari2010mfcc}. Compared to this a CI uses an algorithm that generates electrodograms, containing electrical stimulation information of a maximum of only 22 channels (usually 8 $\sim$ 22) that is almost half the number of filter-banks used for the normal speech recognition process. Therefore our assumption here is that as a lot of speaker-specific feature information would be lost, the speaker-identification capabilities of a CI-user will also deteriorate extensively and even more in noisy conditions.
 
In this study, we focus on the ability of CI users to quantify a specific feature of the human voice, speaker identity. Although a CI user can understand a good percentage of speech (around 30\%) by the movement of the lips, this study is interested to analyze how well a CI-user can identify a distant speaker which include voice from a radio or someone over the phone. To perform this study, CI auditory stimuli, represented as electrodograms, are used as a feature for CI users. Figure~\ref{fig:Electrodogram} represents a electrodogram used to train and test the speaker model in this study. Electrodograms are similar to spectrograms, except electrical stimulation information are simulated and presented as a function of time.

\def\factor{0.9}
\begin{figure}[th]
  \centering
  \includegraphics[width=\factor\linewidth]{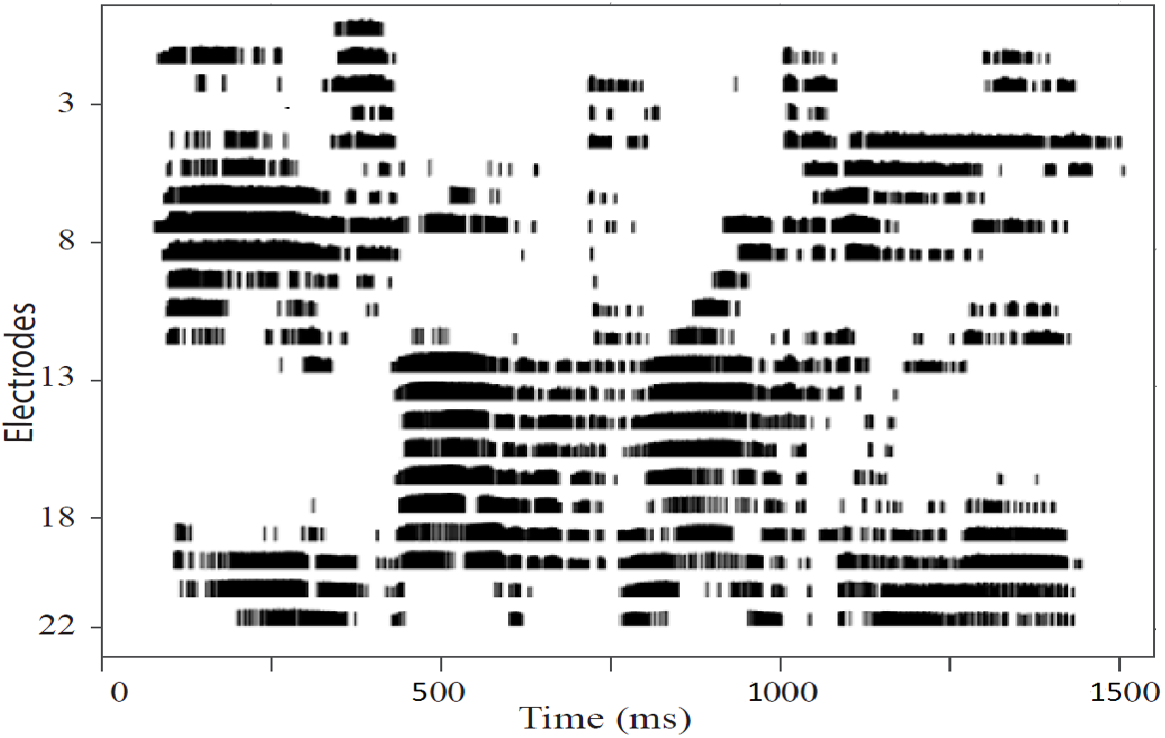}
  \caption{Cochlear Implant electrode stimulation response as an electrodogram}
  \vspace{-10pt}
  \label{fig:Electrodogram}
\end{figure}

Electrodograms reflect the CI auditory stimulation as acoustic time versus frequency-to-electrode allocation \cite{teoh2003acoustic, Mamun2019CNN}. I-Vector features are extracted from the electrodograms and used to train a Gaussian mixture model- universal background model (GMM-UBM) and probabilistic linear discriminant analysis (PLDA) based speaker model. An in-set data of unlabeled speakers is then used in the testing phase under quiet and noisy environments. 

The paper is organized as follows, Sec. 2 briefly explains our proposed speaker identification system. The results under quiet and noisy conditions are presented in Sec. 3 followed by acknowledgment and conclusion.

\section{Methodology}
\begin{figure}[t]
  \centering
  \includegraphics[width=\linewidth]{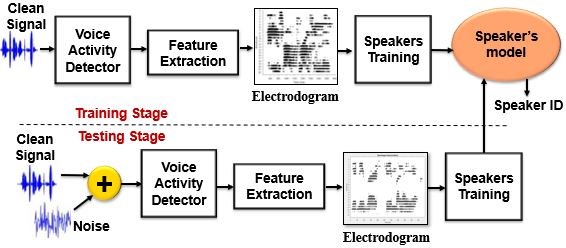}
  \caption{Block diagram of proposed Speaker Identification system using CI auditory stimuli features.}
\vspace{-10pt}
  \label{fig:proposed_system}
\end{figure}

This section briefly explains the proposed method to quantify speaker identification capability of CI-users. The basic block diagram of the proposed system is depicted in Fig.~\ref{fig:proposed_system} . Input speech signal is sent to voice activity detector (VAD) to remove the silent period. The processed signal is then fed to the CI signal processing encoder \cite{clark1986university, arndt1999within}, to generate electrodograms for different speech tokens from each speaker. The well-known GMM-UBM and PLDA classifier are used to train the electrodograms for each speaker. Finally, features extracted from test speaker are then used to identify the correct speaker model using maximum likelihood function, and thereby identify the correct speaker. The performance of the proposed system is evaluated under quiet and noisy conditions.

\subsection{Pre-processing}
Clean signals are submitted to the VAD algorithm to remove the unnecessary silent periods of the signal \cite{brookes1997voicebox}. It also detects the unvoiced part of the signal and removes it to provide the voiced part of the signal as an output of the pre-processor.

\subsection{Feature Extraction}
In this study, electrodograms are used as the auditory stimuli feature for CI system. Electrodogram (shown in Fig.~\ref{fig:Electrodogram}) is a 2-D time-electrode representation which is constructed by combining current levels from 22 electrodes. ACE strategy is used to simulate the received signal and generate electrodograms from speech data. ACE is designed to customize sounds by combining the benefits of pitch information of the SPEAK \cite{skinner2002speech} strategy, with the higher rates of simulation offered by the CIS strategy. The result is an advanced strategy that can be customized to meet each CI-user’s hearing needs.

\begin{figure}[th]
  \centering
  \includegraphics[width=\linewidth]{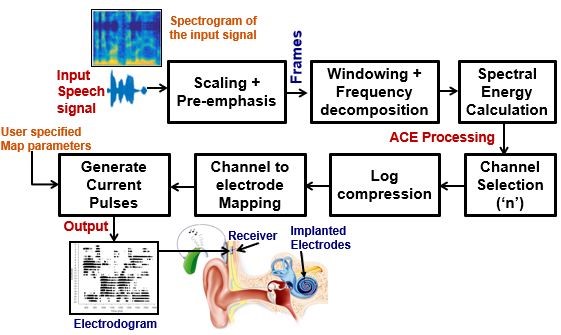}
  \caption{Basic block diagram of ACE processing strategy used in this study to simulate the CI-users signal.}
  \vspace{-10pt}
  \label{fig:ACE}
\end{figure}

The basic block diagram of the feature extraction using ACE processing strategy is shown in Fig.~\ref{fig:ACE} \cite{embc2019}. The incoming signal is pre-processed to emphasize the higher frequency components of the signal. The pre-emphasize signals then divided into frames using a Hamming window (87.5\% overlap between adjacent frames) of length of 128 samples (10 ms) and the envelope (ENV) of each frame calculated. The ENV of each frame is passed through a 22 band pass filter-banks with center frequency specified by Cochlear Corporation. The individual CI-user’s mapping function is then applied on the selected bands and finally electrical pulses are generated for a set of speaker dependent map parameters \cite{vongphoe2005speaker, wilkinson2013voice}. The electrodograms are then generated for each speech signal using electrical pulses.

\subsection{Speaker Training Model}
\subsubsection{GMM-UBM Classifier}
The standard training method for GMM models uses maximum a-posteriori (MAP) adaptation of the means of the mixture components based on speech from a target speaker. To compensate for speaker and channel variability, we stack the means of the GMM model to form a GMM mean super vector \cite{campbell2006support} which has been highlighted in recent research. GMM parameters are iteratively refined by the Expectation Maximization (EM) algorithm that monotonically increases the likelihood of the estimated model for the observed feature vectors \cite{bilmes1998gentle}, and the estimated parameters can be adapted to the new data by MAP adaptation. The GMM speaker model is adapted with the UBM-based training data of each speaker's data to make the system faster, stable, and having better performance \cite{reynolds2000speaker}. In this study, a GMM-UBM classifier with 512 mixture components is used to train the proposed features to generate a model for each speaker.  

\subsubsection{GMM-UBM-I-vectors-PLDA}
\begin{figure}[th]
  \centering
  \includegraphics[width=\linewidth]{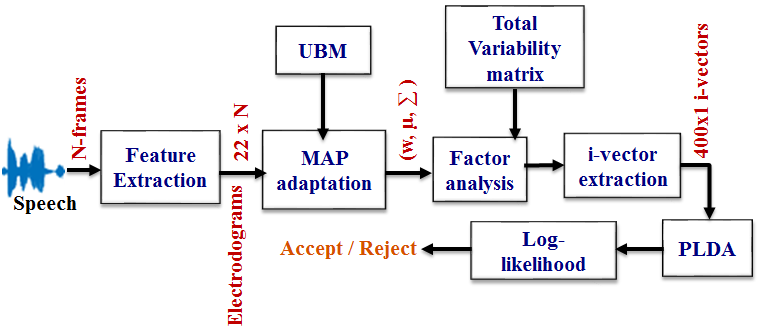}
  \caption{Basic block diagram of speaker traning model using I-Vector-PLDA system.}
  \vspace{-10pt}
  \label{fig:PLDA}
\end{figure}

I-Vector is a low dimensional vector containing both speaker and channel information acquired from a speech segment. PLDA, which is closely related to joint factor analysis (JFA) used for speaker recognition, is a probabilistic extension of linear discriminant analysis. Unlike conventional GMMs, where the correlations are weakly modelled using the diagonal co-variance matrices, PLDA captures the correlation of the feature vectors in subspaces without vastly expanding the model. When PLDA is used on an i-Vector, dimension reduction is performed twice: first in the i-vector extraction process and second in the PLDA model. Figure~\ref{fig:PLDA} represents the speaker modeling system using i-Vector combined with probabilistic linear discriminant analysis (PLDA).

\subsection{Text Corpora}
\subsubsection{Text-Dependent Database}
A small dedicated ‘University of Malaya’ database as a text dependent database is used which consists of 390 signals collected from 39 speakers (25 males and 14 females) \cite{mamun2014robust}. Audio signals were recorded in a noiseless room with a sample rate of 8 kHz where each speaker uttered ‘University Malaya’ 10 times in different sessions. For this study, 70\% (clean) of recorded random data from each speaker is used for training and the remaining 30\% is used to test performance of the proposed system.

\subsubsection{Text-Independent Database}
Next, more extensive corpora based on SRE-18 and YOHO-database are used to test the CI-user ability with text-independent database. This study used 50 speakers with 60 samples ($\sim$ 8 second each) per speaker from both SRE-18 and YOHO database. SRE-18 database \cite{doddington2000nist} is a dataset provided by the National Institute of Standards and Technology for the Speaker Recognition Evaluations (SRE) series conducted by NIST since 1996. The SRE-18 database is composed of telephone conversations collected outside North America, Voice over IP (VOIP) data and audio from Video (AfV). The proposed system is also evaluated with the speech signal from YOHO database. YOHO database is a large scale high quality dataset collected by the ITT (Inter Telephone Telegraph) technical institute in 1989 and is frequently used for speaker identification and verification systems \cite{campbell1994yoho}. Each speaker has four sets of enrollment sessions with 24 independent utterances (with three two-digit number, e.g., 27-82-39) for each enrollment session.
 
 The sample rate for each dataset is 8 kHz. For this study, seventy five percent of speech tokens from each speaker were randomly selected for training the speaker model, and the remaining are used for testing the speaker model.

\section{Results}
This section represents the CI user’s performance to quantify speaker identity for both text dependent and independent databases. The performance of the system was calculated for three different databases, with each experiment repeated 10 times and average scores reported in this section. The effects of GMM parameters are also explained for further analysis.

\subsection{SID performance with Text-dependent database}
\subsubsection{Speaker-identification under quite conditions}
Speaker identification performance of CI users were predicted for text dependent database based on CI auditory stimuli derived from electrodograms. The accuracy was evaluated using two different classifier system (i.e., GMM adapted with UBM and i-Vector-PLDA). Performance was calculated for 39 speakers from “UM database” used in the training session in quiet environment. The result shows that, CI user auditory stimuli based system can effectively predict the speaker identity when text is common. The highest accuracy was 95\% and 99\% for 39 speakers using GMM-UBM and i-Vector-PLDA based classifier, respectively. Therefore, it is expected that CI users would have high speaker identification ability for text dependent database.
\begin{table}[th]
  \centering
  \caption{Speaker Identification accuracy for different GMM mixtures (\%): Text-dependent Corpus.}
 \includegraphics[width=\linewidth]{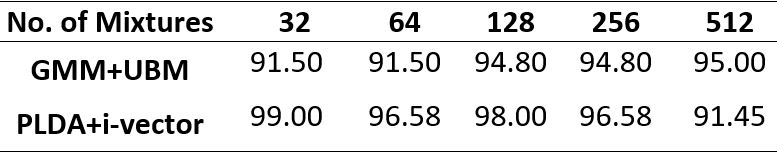}
 \end{table}
 \vspace{-10pt}

\subsubsection{Effects of GMM parameter on SID performance}
Table 1 illustrates the accuracy of the system for increasing number of acoustic distribution of components for the GMM. The system was trained and tested for different speech token from text dependent database for 10 dB speech shaped noise. The accuracy of the system was then evaluated for increasing number of GMM parameters. Increased accuracy is achieved when the number of GMM components is expanded (GMM-UBM based method). However, i-vector based method showed the opposite trend. This could be due to the overestimation speaker information. Moreover, with an increased GMM distribution, the computational time for log-likelihood iterations also increases which is needed to achieve the ideal value for convergence for the GMM.     

\subsection{SID performance with text-independent database:}
\subsubsection{Speaker-identification under quite conditions}
It is expected that a CI-user has more difficulty as the number of speakers increase. Table 2 represents the predicted speaker identification ability of CI-users for text-independent SID. The results were evaluated for 50 different speakers from two different databases (SRE-18 and YOHO)The speaker models were constructed and classified using two well-defined classifiers based on GMM-UBM and i-Vector-PLDA. To evaluate the performance against different number of speakers, the model was trained and tested for 4, 12, 24, 36, and 50 speakers. To check the consistency, the experiments were repeated 10 times and average results are presented in the table.  In general, a CI user auditory stimuli based system can predict speaker identity under quiet conditions for both databases. The system can effectively identify the speaker’s identity when tested with 4-12 speakers. However, their performance decreases as more and more speakers are incorporated into the training set. Performance of the system is higher when the PLDA based classifier is used versus that for the GMM-UBM based classifier. 
\begin{table}[th]
  \centering
  \caption{Closed-set Speaker Identification accuracy for text independent data in percentage (\%).}
 \includegraphics[width=\linewidth]{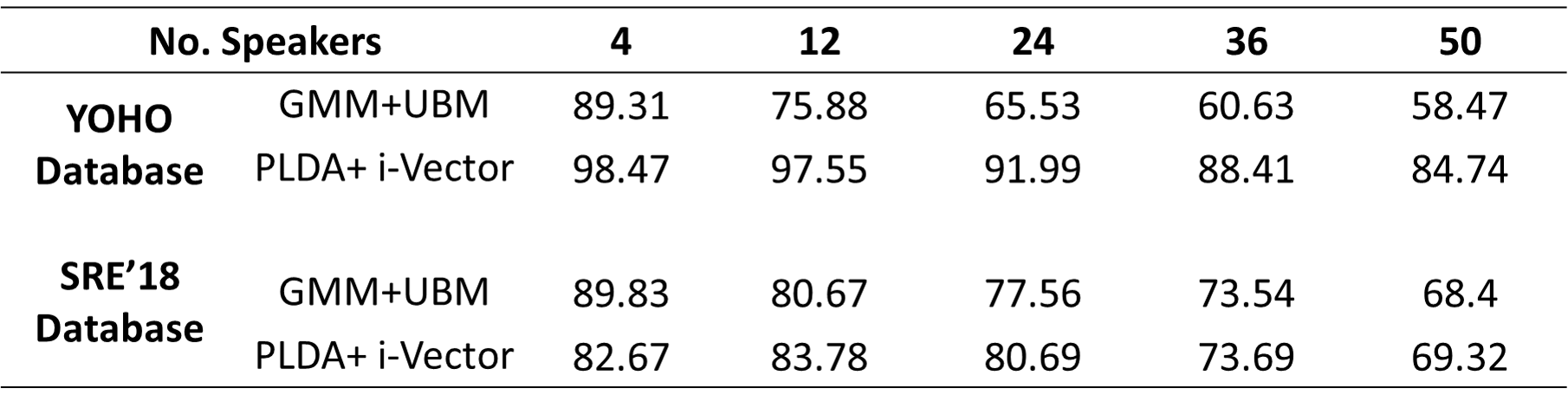}
 \end{table}
   \vspace{-10pt}

\subsubsection{Speaker-identification accuracy under noise conditions}
To assess CI-users’ performance of speaker identification under noisy conditions, system performance was evaluated by adding 10dB noise. The speaker model was trained with clean data from both the YOHO and SRE databases and then tested for speech tokens contaminated with 10 dB White Gaussian noise (WGN) and Speech Shaped noise (SSN). System performance was also evaluated for varying number of speakers (4, 12, 24, 36, and 50). Training and test samples were randomly selected, and the experiment repeated 10 times to reduce any system bias. Finally, the average results of the experiments are reported in Table 3.

\begin{table}[t]
  \centering
  \caption{Speaker Identification accuracy under noisy conditions in percentage (\%).}
 \includegraphics[width=\linewidth]{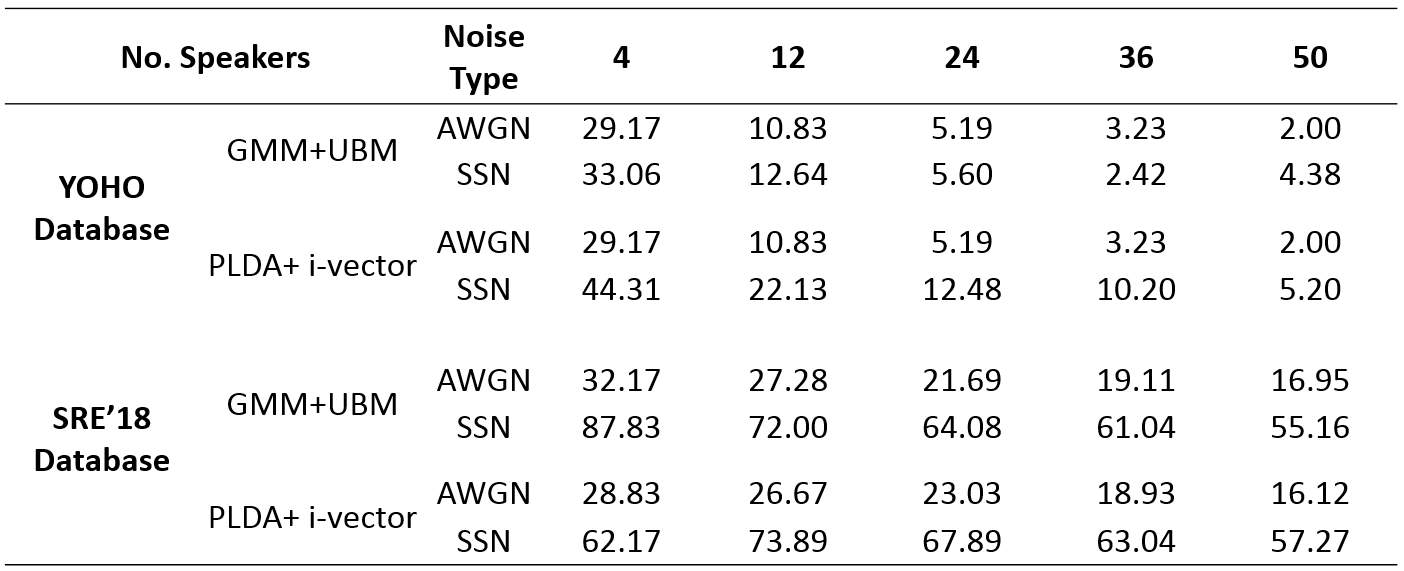}
 \end{table}
   \vspace{-10pt}

Although, speaker identification accuracy is high in quiet environment, it drastically falls in noisy conditions. Moreover, the proposed system has higher accuracy for SRE versus the longer test durations in YOHO database, as it contains complete sentences which suggests that the subjects obtain sufficient cues regarding speaker identity. In addition, system performance using PLDA classifier is respectively higher than for GMM-UBM under noisy conditions. Therefore, it is clear that the speaker identification accuracy is very poor (approximately zero) under noise which reflects the reality of CI-user capability.


\section{Conclusion}

This study quantifies the speaker identification capability for CI-user based on a parameterized electrodogram feature set. Electrodograms were generated using ACE signal processing strategy for speech signal from three different databases. To quantify the performance of speaker ID for normal hearing (NH) subjects versus cochlear implant (CI) subjects, two alternate time-frequency acoustic front-ends were considered to represent NH versus CI based human SID performance. Two different backend classifiers were used- GMM-UBM and i-Vector–PLDA along with GMM to evaluate the CI-user performance. The results showed that, the CI-based auditory stimuli (e.g., parameterized electrodograms) is effective for speaker ID under quite conditions (e.g., high accuracy of 90~99\%). It is also shown that, the CI based acoustic representation within the i-Vector based speaker ID system is more successful (98\%) vs. the GMM-UBM based system (94\%). However, CI electrodogram based SID results are completely confused and unable to predict speakers under noisy conditions, suggesting that CI-user auditory stimuli is not capable of representing speaker ID traits for CI listeners. An important analysis is that CI-users can easily predict speakers identity when text was fixed, but deteriorates for text independent scenarios. For future work, it is suggested that a parallel investigation using CI-users for a subjective study could validate these corresponding MFCC (NH) and Electrodogram (CI) based SID systems. Finally, it is suggested that the resulting proposed systems could be applied to improve the signal processing strategies in cochlear implant processors to improve speaker characterization for CI listeners in both quiet and noisy environments, thereby improving quality-of-life experience for CI users. 

\section{Acknowledgement}
This work was supported primarily by Grant No. R01 DC016839-02 from National Institutes of Health (NIDCD); and partially by University of Texas at Dallas from the Distinguished University Chair in Telecommunications Engineering held by J.H.L. Hansen.

\bibliographystyle{IEEEtran}
\bibliographystyle{IEEEtran}
\bibliography{mybib}

\end{document}